\documentclass[runningheads]{cl2emult}

\usepackage{makeidx}  % allows index generation
\usepackage{graphicx} % standard LaTeX graphics tool
\usepackage{subeqnar} % subnumbers individual equations
\usepackage{multicol} % used for the two-column index
\usepackage{cropmark} % cropmarks for pages without
\usepackage{eso}      % placeholder for figures
\makeindex            % used for the subject index
\begin{document}
\title*{SDSS-RASS: Next Generation of Cluster--Finding Algorithms}
\toctitle{SDSS-RASS: Next Generation of Cluster--Finding Algorithms}
\titlerunning{SDSS-RASS: Next Generation of Cluster--Finding Algorithms}
\author{Robert Nichol\inst{1}
\and Chris Miller\inst{1}
\and Andy Connolly\inst{2}
\and Shan-Shong Chong\inst{1}
\and Chris Genovese\inst{1}
\and Andrew Moore\inst{1}
\and Daniel Reichart\inst{1}
\and Jeff Schneider\inst{1}
\and Larry Wasserman\inst{1}
\and Jim Annis\inst{3}
\and John Brinkman\inst{4}
\and Hans Bohringer\inst{5}
\and Francisco Castander\inst{6}
\and Rita Kim\inst{7}
\and Tim McKay\inst{8}
\and Marc Postman\inst{9}
\and Erin Sheldon\inst{8}
\and Istvan Szapudi\inst{10}
\and Kathy Romer\inst{1}
\and Wolfgang Voges\inst{5}
}
\authorrunning{Nichol et al.}
% if there are more than two authors,
% please abbreviate author list for running head
%
%
\institute{Carnegie Mellon Univ., 5000 Forbes Ave., Pittsburgh,PA-15217
\and Dept. of Physics and Astronomy, Univ. of Pittsburgh, Pittsburgh,
PA-15260 
\and Fermilab, P.O. Box 500,
Batavia, IL 60510 
\and Apache Point Obs., P.O. Box 59,
Sunspot, NM 88349-0059
\and Max Planck Institute for Extraterrestrial Physics,
85748 Garching, Germany                            
\and Observatoire Midi-Pyr\'en\'ees, 14 ave Edouard Belin,
Toulouse, F-31400, France
\and Princeton University, Dept. of Astrophysical Sciences, Peyton Hall, Princeton, NJ 08544
\and Univ. of Michigan, Dept. of Physics,
	500 East University, Ann Arbor, MI 48109
\and STScI, 3700 San Martin Drive, Baltimore, MD-21218
\and CITA,
Univ. of Toronto,
Toronto, Ontario, M5S 3H8, Canada
}

\maketitle              

\begin{abstract}
\index{abstract} We outline here the next generation of cluster--finding
algorithms. We show how advances in Computer Science and Statistics have
helped develop robust, fast algorithms for finding clusters of galaxies in
large multi--dimensional astronomical databases like the Sloan Digital Sky
Survey (SDSS). Specifically, this paper presents four new advances: (1) A new
semi-parametric algorithm -- nicknamed ``C4'' -- for jointly finding clusters
of galaxies in the SDSS and ROSAT All--Sky Survey databases; (2) The
introduction of the {\it False Discovery Rate} into Astronomy; (3) The role of
kernel shape in optimizing cluster detection; (4) A new determination of the
X--ray Cluster Luminosity Function which has bearing on the existence of a
``deficit'' of high redshift, high luminosity clusters. This research is part
of our ``Computational AstroStatistics'' collaboration (see Nichol et
al. 2000) and the algorithms and techniques discussed herein will form part of
the ``Virtual Observatory'' analysis toolkit.
\end{abstract}

\section{Introduction}

Clusters of galaxies are critical cosmological probes for two fundamental
reasons. First, galaxy clusters are the most massive virialised objects in the
universe and reside in the tail of the mass distribution function. Thus the
distribution, density and properties of clusters are very sensitive to the
mean mass density of the universe. For example, in a low $\Omega_m$ universe,
the evolution of the mass function terminates at early epochs, thus
``freezing'' the number of massive systems. The opposite is
true for $\Omega_m=1$, were the mass function continues to evolve
up to the present epoch (see Press \& Schtecter 1974; PS).  Unlike other
cluster methods of measuring $\Omega_m$ (mass-to--light ratios, baryon
fractions {\it etc}), this technique provides a global measure of $\Omega_m$
and is relatively insensitive to $\Omega_{\Lambda}$ (see Romer et al. 2000).
For a more complete description of PS and this effect, the reader is referred
to Viana \& Liddle (1999), Reichart et al. (1999) \& Borgani et al. (1999),
and references therein.  Many authors have used this technique with a
preponderance of the evidence for a low $\Omega_m$, but the observed scatter
between the various measurements is large.  For example, we have seen
$\Omega_m=0.3\pm0.1$ (Bahcall et al. 1997), $\Omega_m=0.45\pm0.2$ (Eke et al.
1998), $\Omega_m=0.5\pm0.14$ (Henry 1997), $\Omega_m\sim0.75$ (Viana \& Liddle
1999), $\Omega_m\sim0.85\pm0.2$ (Sadat et al. 1998), and
$\Omega=0.96^{+0.36}_{-0.32}$ (Reichart et al. 1999). This scatter could be the
combination of several effects including small sample sizes, errors in the
survey selection functions as well as the necessity to compare local samples
of clusters to more distant samples in any effort to see an evolutionary signal.

Secondly, each cluster represents a sample of galaxies that has formed at
roughly the same time under roughly the same initial conditions. Thus, each
cluster is a laboratory for understanding how galaxies form, evolve and
interact with their environment. Such phenomena are usually known as the {\em
morphology-density relation} and/or the {\em Butcher--Oemler Effect}.  There
are many proposed models to explain these phenomena including ram--pressure
stripping, pressure--induced star formation, temperature--inhibited star
formation, galaxy harassment, shocks from cluster merger events {\it etc.}
As yet, we do not have a definitive answer but observations of clusters - as a
function of the cluster properties - will help to dis--entangle these
different possible mechanisms.

It is clear that we need larger, more objective, cluster catalogs to help
obtain high precision measurements of $\Omega_m$ as well as determine the
physical mechanism(s) behind galaxy evolution in clusters (see Kron 1993 for a
discussion of the role selection effects could play in the Butcher--Oemler
Effect).  To significantly move beyond previous work, any new sample of
clusters should possess the following key attributes: {\it i)} Contain
many thousands of clusters; {\it ii)} Possess a well understood selection
function; {\it iii)} Probe a broad range in redshift and mass, and {\it iv)}
Possess physically meaningful cluster parameters.  There are several cluster
projects under-way that satisfy these criteria (see, for example, Collins et
al. 2000; Romer et al. 2000; Gladder \& Yee 2000; Ebeling et al. 2000)
including the SDSS--RASS Cluster (SRC) survey which we discuss below.

\section{The SDSS--RASS Cluster Catalog}

The SRC is a new sample of clusters constructed through the unique union of
the Sloan Digital Sky Survey (SDSS; see York et al. 2000 \& Castander et al. 2000) and the ROSAT All--Sky Survey (RASS; see Voges et al. 2000).
Using state--of--the--art cluster--finding methodologies, we plan to construct
a sample of $\sim50,000$ clusters and groups that will span a large dynamic
range in both redshift and mass as well as possessing a well--understood
selection function. Moreover, we will endeavor to measure physically
meaningful parameters for these clusters like X--ray and optical luminosities,
lensing masses, velocity dispersions {\it etc.}  Finally, the SRC also
represents a critical first step in the construction of the ``Virtual
Observatory'' (VO); for the first time, we plan to perform a joint
optical--X-ray cluster selection using these existing, multi--dimensional, data
archives.

\begin{figure}[t]
\sidecaption\includegraphics[width=.6\textwidth]{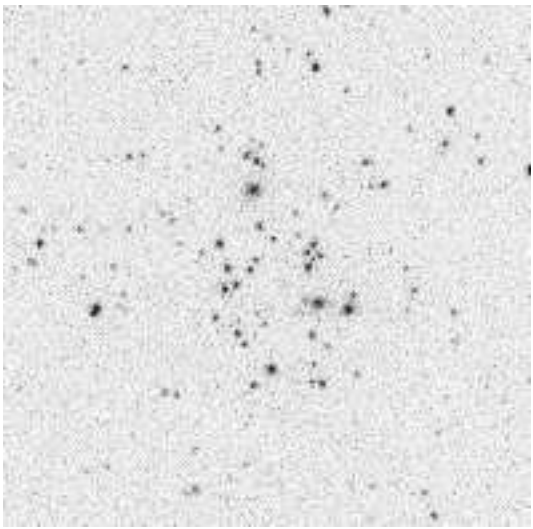}
\caption[width=.4\textwidth]{A greyscale SDSS image of RXJ0254, an X--ray
luminous cluster ($L_x(44)=4$) at $z=0.36$.  The brightest member of this
cluster is $r'=18.7$ and is just below the main SDSS spectroscopic galaxy
target limit. However, this galaxy has the correct colors and magnitude to be
included in the SDSS Bright Red Galaxy (York et al. 2000) sample.  The
signal--to--noise of the individual galaxies in this image is reasonably high
($>20$) thus allowing us to study their characteristics.
\label{SHARC}
}
\end{figure}

To illustrate the power of combining the SDSS and RASS, we have constructed a
preliminary SRC catalog which was constructed simply from the
cross--correlation of the RASS source lists (both bright and faint source
catalogs from Voges et al. 2000) with a preliminary SDSS cluster sample
constructed by Annis et al. (2000) using the maxBCG algorithm. We discuss
our new method of finding clusters in the SDSS--RASS in Section \ref{method}

\begin{figure}[t]
\sidecaption\includegraphics[height=3.0in,width=.7\textwidth]{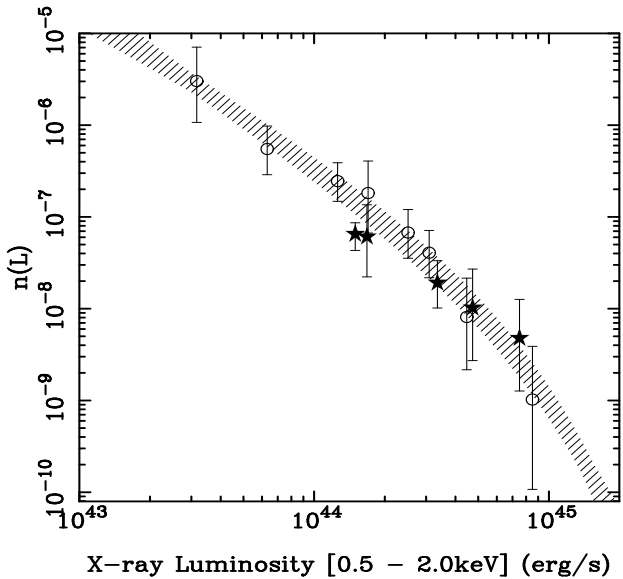}
\caption[width=.3\textwidth]{We show here a preliminary XCLF for the SRC survey
($0.0 < z < 0.45$; stars). The shaded region is the local ($0.0 < z < 0.3$) x-ray luminosity
function, plus $1\sigma$ errors, as measured by de Grandi et al (1999) and
Ebeling et al (1998).  The circles are the SHARC ($0.3 < z < 0.7$ Burke et
al. 1997; Nichol et al. 1999).
Poisson errors are shown.
\label{lf}
}
\end{figure}

In Figure \ref{SHARC}, we show an example of a high redshift cluster detected
in this preliminary catalog, while in Figure \ref{lf}, we show a very
preliminary determination of the SRC X--ray Cluster Luminosity Function (XCLF)
(21 clusters above $L_{\rm ROSAT}(44)>1$ in the redshift range
$0.0<z<0.45$). Here we have used runs 752 \& 756 ($213\, {\rm deg^2}$) and 94
\& 125 ($90\, {\rm deg^2}$) from the SDSS data (see York et al. 2000) which we
have naively assumed to be complete to a flux limit of $4$ and $8\times 10^{-13}\, {\rm erg/s/cm^2}$
respectively. This assumption is valid for our luminous clusters but breaks
down for lower luminous clusters and undoubtedly accounts for the ``low'' data
points in our XCLF at $L_{\rm ROSAT}(44)\sim1$.  We have also conservatively
cut the catalog at $z<0.45$ and removed suspicious
match--ups by hand (see Sheldon et al. 2000).  
However, the SRC does have candidate
$z\simeq0.7$ clusters which will require spectroscopic confirmation (Annis et
al. 2000). 

We present this data to illustrate the power of the SRC for
detecting real clusters to high redshift; the combination of these two, large
area, catalogs (SDSS and RASS) allows us to push to fainter fluxes (and thus
higher redshift) than one would be able with either survey on its
own. Finally, the XCLF presented here is fully consistent with other
measurements and adds new information regarding the existence of a ``deficit''
of high redshift, high luminosity clusters which has been debated by many
authors ({\it e.g.}  Reichart et al. 1999; Nichol et al. 1999; Gioia et
al. 1999).  This is because the SRC covers a large area of sky ($\simeq300{\rm
deg^2}$ at present) which is vital for detecting massive, X--ray bright
clusters at high redshift (see Ebeling et al. 2000).

\section{Finding Clusters in Multiple Dimensions}
\label{method}

Over the last few years, there has been significant progress in the
development of new cluster--finding algorithms primarily driven by the quality
and quantity of new data as well as the increased availability of fast
computing. These new methods include the matched--filter algorithm (Postman et
al. 1996, Kawasaki et al 1998, Kepner et al 1999, Bramel et al. 2000), the wavelet--filter (Slezak et al 1990), the
``photometric redshift'' method (Kodama et al. 1999), 
and the ``density--morphology'' relationship (Ostrander
et al. 1998) or the E/S0 ridge--line (Gladders \& Yee 2000). 

Here we outline a new algorithm we have developed which exploits the quality
and quantity of data available to us. For instance, the SDSS will provide
accurate, calibrated magnitudes in 5 filters ($u^{'} g^{'}, r^{'}, i^{'},$ and
$z^{'}$) as well as accurate star-galaxy separation, galaxy shapes,
photometric redshifts and estimates of the galaxy type. These increased number
of observables allow us to refine our definition of a cluster to be an
overdensity of galaxies in both space (on the sky and in redshift) and
rest--frame color. Furthermore, we can require that the cluster members be
early--type galaxies and that the cluster, as a whole, is coincident with
extended hard X-ray emission.  The motivation for this definition of a cluster
is the growing body of evidence that the cores of clusters are dominated by
ellipticals of the same colors suggesting they are coeval (see Gladders \& Yee
2000), and possess a hot, intracluster medium (see Holden et al. 2000).  Via
this definition, we can radically increase the signal--to--noise of clusters
in this multi--dimensional space thus effectively removing projection effects
and X--ray mis--identifications which presently plague optical and X--ray
cluster surveys respectively.  We have nicknamed this algorithm ``C4'' for
4--color clustering.

We note that this definition may bias us against certain types of clusters
{\it e.g.} young systems where the X--ray gas may be more diffuse (thus a
lower emissivity) and/or have a bluer, less homogeneous, galaxy
population. However, over the redshift range probed by SDSS \& RASS ($z<0.8$),
most clusters are expected to be well evolved since they have formation epochs
of $z\ge2$. Moreover, we will need to quantify our exact selection function
regardless of the algorithm used (see Section \ref{bias}).

We present here a brief overview of the C4 algorithm and then present specific
details about parts of the algorithm below.  We start by considering galaxy
$X_i$ which is any detected galaxy with a known or photometric redshift. The
fundamental question the C4 cluster-finding algorithm poses is: {\em Is there
an overdensity of cluster-like galaxies about galaxy $X_i$?} Our solution for
answering this question is rather simple. First, we count $N_i$, the number of
galaxies in a multi--dimensional aperture around galaxy $X_i$.  The aperture
(discussed in detail in Section \ref{aperture}) is defined in angular,
redshift and color space. Second, for each test galaxy $X_i$, we measure a
field distribution, $F(X_i)$, which is constructed via Monte Carlo
realisations of placing the same size aperture as the test galaxy on thousands
of randomly chosen galaxies in regions of similar extinction and seeing as
$X_i$.  Third, if galaxy $X_i$ is in a clustered region, then $N_i$ will lie
in the tail of the distribution and we can measure a probability, $p_i$, that
galaxy $X_i$ is a member of this field distribution $F(X_i)$.  We then have to
decide what acceptable cut-off in $p_i$ differentiates between cluster
galaxies and field galaxies (see Section 3.2).  Note that C4 does not find
clusters but detects cluster--like galaxies. It is based on a well-defined
description of the field, which is easily measured from a large-volume survey
such as the SDSS.  Those galaxies that have a low probability of being a field
galaxy are then considered to be cluster members. Finally, we use these
cluster--like galaxies to locate the positions of the actual clusters in the
data.

\begin{figure}[pt]
\centering
\includegraphics[height=3.0in,angle=270.]{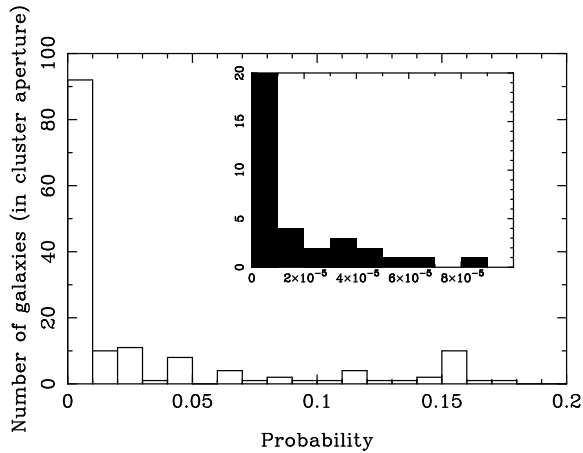}
\caption[width=0.3\textwidth]{The distribution of $p$--values for the 183 galaxies
within the aperture of SRC galaxy RXJ0254 (Figure \ref{SHARC}). Inserted is a
blow--up of small $p$--values which illustrates the strong peak near zero {\it
i.e.} probability in 6--dimensional space that these galaxies were drawn from the
field distribution. This illustrates the strength of clustering one can obtain
since 6--dimensional space is mostly empty and thus finding any grouping
of galaxies on the sky with the same redshift \& color is enough to make it highly unlikely that they are drawn from a random field population.
\label{fdr1}
}
\end{figure}

\subsection{Choosing The Aperture: Shape Doesn't Matter}
\label{aperture}

A critical part of computing $N_i$ -- the galaxy count in multi--dimensions
around our test galaxy -- is the choice of the size and shape of the counting
aperture {\it i.e.} we are using a kernel to smooth the data.  We note
here that we focus on defining the width of this kernel or aperture, rather
than the shape, since it is well--known in the statistical literature that the
choice of bandwidth of a kernel is more important when optimally smoothing
data than the exact shape of that kernel.

To discuss this further, suppose that $X_1, \ldots, X_n$ are independent observations from
a probability density function $f(x)$. A common estimator is the kernel
estimator defined by
$$
\hat{f}(x) = \frac{1}{n} \sum_{i=1}^n \frac{1}{h_n} K\left( \frac{x-X_i}{h_n}\right).
$$
The function $K$ is called the kernel and is usually assumed to satisfy $K(x)
\geq 0$, $\int K(x) dx=1$, $\int x K(x) dx=0$.  For example, the Gaussian
kernel is $K(x) = \{ 2\pi\}^{-1/2} e^{-x^2/2}$.  The number $h_n$ is the
bandwidth and controls the amount of smoothing.  One can see from numerical
experimentation that the choice of $K$ has very little effect on the estimator
$\hat{f}$ but the choice of $h_n$ has a drastic effect.  This can also be
proved mathematically.  For example, consider the integrated means squared
error (IMSE) defined by
$$
IMSE = E \int (f(x) - \hat{f}(x))^2 dx
$$
where $E$ is the average or expectation value.  One can derive an analytic
expression for IMSE and from this expression one sees that $K$ has little
effect on IMSE but $h_n$ has a drastic effect.  Details are given in {\em
Density Estimation for Statistics and Data Analysis} by Silverman (1986).  In
fact, one can show that the optimal kernel, which minimizes IMSE, is given by
$K(x) = (3/4) (1- t^2/5)/\sqrt{5}$ for $|x| < \sqrt{5}$ and 0 otherwise.  This
is called the Epanechnikov kernel.  But the efficiency of other kernels (the
ratio of the IMSEs) is typically near 1.  For example, the Gaussian kernel has
an efficiency of 0.95 compared to an Epanechnikov kernel.  In contrast, the
effect of $h_n$ is dramatic, so; how does one find the optimal bandwidth?  This
is clearly difficult and in statistics, it is usually achieved using
``cross-validation'' where one estimates the $IMSE(h_n)$
and finds $h_n$ to minimize this function (see Silverman 1986).

Therefore, we simply use a top--hat kernel in the C4 algorithm and use the
known observables of the galaxy $X_i$ (redshift and colors), as well as known
physical relationships, to define the aperture size or bandwidth.  We start by
defining an input mass scale to search for clusters.  This allows us to
calculate $r_{200}$ which is converted to an angular aperture ($\Delta\theta$)
as function of cosmology. In this conversion, we use either the known or
photometric redshift (see Connolly et al. 1995). Next we define a redshift
aperture, $\Delta z$, which is based on the expected radial velocity
dispersion for a cluster with $r_{200}$ (determined analytically). This is
also convolved with the error on the observed or photometric $z$.  Finally, we
use a 4-dimensional color-aperture, $\Delta_ic$, for $i=1,4$, where the width
of this aperture is simply the measured errors on the SDSS photometric
magnitudes for galaxy $X_i$.

Once we have defined our aperture, we then simply count the number of
neighboring galaxies within this 6--dimensional aperture. We note that all the
SDSS galaxies will possess errors on the redshift (photometric and
spectroscopic) and colors, so instead of counting each galaxy as a single
delta--function in our count $N_i$, we can treat each galaxy as an error
ellipsoids in 6--dimensional space and compute the amount of overlap between
these ellipsoids and the aperture.  This is non-trivial and therefore, we will
Monte Carlo the effect by computing many $N_i$ around each test galaxy $X_i$
perturbing in each case the surrounding galaxies by their observed
errors. This will result in a distribution of $N_i$ for each galaxy, $f(N_i)$.

The counting queries we have outlined above are computationally intensive and
thus, to make this problem tractable, we will exploit the emerging algorithmic
technology of multi--resolutional KD--trees (see Connolly et al. 2000) which
can scale as $N\,logN$, instead of $N^2$, for range searches like those
discussed herein. We defer discussion of these computational issues to a
forthcoming paper.

Having measured $N_i$, we now need to build the field distribution, $F(X_i)$.
Recall, we want to use the same size aperture as that of galaxy $X_i$.  In
order to determine $F(X_i)$, we will count galaxies around a random
distribution of galaxies that lie in regions of similar extinction and seeing
to that of galaxy $X_i$. As above, these field counts can be corrected for
observational errors and thus $F(X_i)$ will be the sum of many count
distributions around each randomly selected field galaxy.

\begin{figure}[pt]
\centering
\includegraphics[height=3.in]{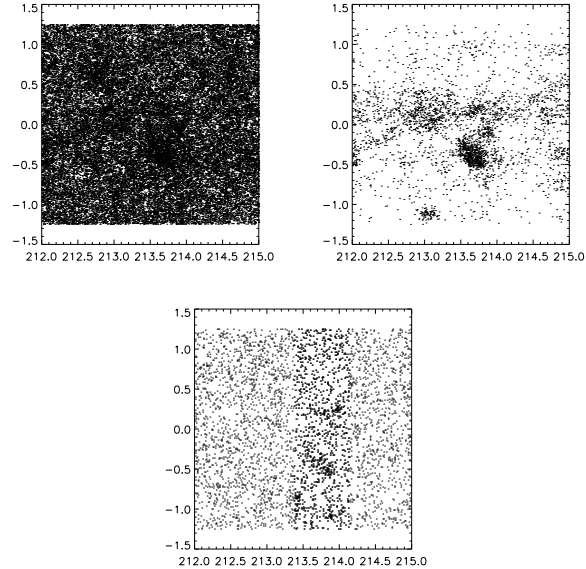}
\caption[]{
(Top left) A $3^{\circ}\times 3^{\circ}$ area from SDSS photometric data 
runs 752 \& 756 (rerun 1). We show all galaxies regardless of redshift, color
or magnitude. (Top right) This is the result of running the C4 algorithm 
on this data using a 5\% FDR threshold. We rejected 4\% of the galaxies
on the left as being field--like galaxies. (Bottom) Same area taken from the RASS hard photon data (each dot is at least one photon). The large cluster 
seen in the field is Abell 1882 (richness class 3 and $z=0.137$). 
The axes
are RA and DEC
\label{chris}
}
\end{figure}

\subsection{Choosing The Threshold: False Discovery Rate}
Once we have $N_i$ and $F(X_i)$ for each test galaxy $X_i$, we can compute the
probability, or $p$--value, that this test galaxy was drawn from the field.
This is achieved by fitting a Poisson distribution to the lower end of the
$F(X_i)$ distribution (which must be the field population) and comparing $N_i$
to that fitted distribution, see Figure \ref{fdr1}.

In this process, we can make two types of errors: (1) falsely identifying a
real field galaxy as cluster-like (false rejection); (2) falsely identfiying a
real cluster galaxy as field like (false non-rejection).  The next critical
decision is to determine the $p$-cutoff below which a galaxy is rejected as
being field--like.

This threshold could be chosen arbitrarily. For instance, for each test we
could apply a $2\sigma$ confidence requirement and reject any galaxy with $p <
0.05$.  However, after $N_{total galaxies} = 1 million$ tests, we would expect
to have made as many as 50000 mistakes through false rejections. This is far
too many mistakes which could be reduced by applying a higher confidence
requirement {\it e.g.}  $4\sigma \rightarrow p_{cutoff} = 3 \times 10^{-7}$.
This leads to the traditional approach of permitting {\it no} false rejections
with 95\% confidence through lowering the $p$-cutoff to $0.05/N_{gal}$ where
$N_{gal}$ is the number of test galaxies.  This is known as the Bonferoni
Method where each individual test is very conservative to allow for {\it no}
false rejections. The disadvantage of this approach is that a lot of
cluster--like galaxies would be mis-classified as field--like ({\it i.e.}
false non-rejections) since we have enforced a very strict limit on the number
of field--like galaxies that are allowed to be mis--classified as
cluster--like galaxies. This is an extreme case where one sacrifices errors in
one direction for the control of mis--classification errors in the opposite
direction.  If there were no cluster--like galaxies in our test sample of
galaxies, then the Bonferoni Method would be correct, however $\sim20$\% of
all galaxies live in cluster \& group environments, so if we using this method
we would loose significant sensitivity to detecting these galaxies.

Instead of either of the above two thresholding techniques,
we use the newly devised {\em False Discovery Rate} (FDR; Abramovich et
al. 2000).  
%To see the power of FDR, consider we want to ensure, on average, a
%$>1$\% probability of incorrectly rejecting our test galaxies, $X_i$, as
%field--like galaxies {\it i.e.} we only want 1\% of our cluster--like galaxies
%to be in error.
FDR is a new, more
adaptive approach which, in the limit of all the galaxies in the test sample
being field--like, would be equivalent to the Bonferoni Method. However, FDR
becomes less conservative, and makes fewer errors in the other direction, as
we diverge from this idealized case.
In practical terms, FDR allows us to define in advance a desired false
detection rate {\it i.e.} up--front only $\alpha\times 100\%$ of the rejected
galaxies are in error based on our null
hypothesis. Moreover, the FDR procedure is simple:
\begin{enumerate}
\item{For each test galaxy, calculate a $p$-value based on the null hypothesis that it is a field galaxy.}
\item{Sort these according to increasing $p$-value.}
\item{Rank the $p$-values as a function of $n/N$ where $n$ is the
$n^{th}$ galaxy's $p$-value out of $N$ total test galaxies.}
\item{Draw a line with slope ($\alpha$) and intercept 0.}
\item{From the right, determine the {\it first crossing} of the line with
the ranked $p$-values.}
\item{Anything with a $p$-value smaller than the crossing $p$-value is
rejected.}
\item{For our cluster--finding algorithm, these rejected galaxies are our
cluster--like galaxies, with at most $\alpha\times 100\%$ errors based on our
null field galaxy hypothesis.}
\end{enumerate}
The beauty of FDR is that {\it (a)} it is simple, {\it (b)} it possesses a
rigorous statistical proof, and {\it (c)} it works for highly correlated
data (in this case the slope becomes $\alpha/{\rm log}N$). In summary, FDR is a new tool in statistics which has the potential to
significantly enhance astronomical analyses; this is the first application of
this new statistic in astronomy and demonstrates the power of our
``Computational AstroStatistics'' collaboration.  Other possible applications
could be determining the sky threshold values in source detection, point--source
extraction in CMB analyses {\it etc.} We will explore these application in a
forthcoming paper.

\subsection{Clustering Cluster--like Galaxies}

In Figure \ref{chris}, we show the results of running our C4 algorithm, with
FDR, on the SDSS commissioning photometric data (York et al. 2000).  The next
task is to ``cluster'' these cluster--like galaxies into a sample of unique
clusters of galaxies.  To do this, we employ the Expectation Maximization (EM)
algorithm which is a mixture model of Gaussians designed to be highly adaptive
and multi--resolutional in nature.  Moreover, it is naturally
multi--dimensional thus allowing us to feed it three galaxy spatial
coordinates (angular position and redshift) as well as three X--ray dimensions
(hard photon energy and angular position). The rational for jointly clustering
the X--ray and optical data is to drastically increase the signal--to--noise
of distant and poor clusters by suppressing projection effects and X--ray
mis--identifications. The details of mixture--model clustering and the EM
algorithm can be found in Connolly et al. (2000). In addition, to using EM we
plan to investigate adaptive kernel density estimators.  We note here that
during the clustering process additional information can be used to ``mark''
or ``up--weight'' galaxies. For example, SDSS data can be used to determine if
a galaxy is elliptical--like based on the imaging morphology (likelihood fits
to each galaxy for both a da Vaucouleurs and exponential profile), photometric
spectral classification information of Connolly \& Szalay (1999) as well as
spectral classifications where available (see Castander et al. 2000 for
details).

\subsection{The Selection Function \& Systematic Biases}
\label{bias}

The most direct method of
quantifying the selection function of our SDSS cluster catalog is via Monte
Carlo simulations (see Bramel et al. 2000; Kim et al.2000).  These simulations
can then be converted to an effective area, or volume, of the survey as a
function of the input cluster parameters (redshift, luminosity {\it etc}).  In
addition to simulations, we can obtain information about the selection
function of the various SDSS cluster catalogs (Annis et al. 2000; Kim et
al. 2000) by cross-correlating them against each other and the low redshift
SDSS galaxy redshift survey. These different SDSS cluster catalogs use
different cluster selection criteria since they are explore different science
issues and are therefore, complementary.

In summary, we have outline here our new C4 cluster--finding algorithm that
exploits the quality and quantity of the multi--dimensional survey data now
becoming available. It also exploits new techniques and algorithms coming out
of Computer Science and Statistics. We plan to jointly ``cluster'' optical and
X--ray data to help improve the signal--to--noise of distant and/or poor
clusters/groups of galaxies. This represents a first step toward the ``Virtual
Observatory''; the joint scientific analysis of archival multi--wavelength
survey databases. The algorithms and software we develop will become part of
the ``Virtual Observatory'' analysis toolkit.

%INDEX%%%%%%%%%%%%%%%%%%%%%%%%%%%%%%%%%%%%%%%%%%%%%%%%%%%%%%%%%%%%%%%
\clearpage
\addcontentsline{toc}{section}{Index}
\flushbottom
\printindex
%%%%%%%%%%%%%%%%%%%%%%%%%%%%%%%%%%%%%%%%%%%%%%%%%%%%%%%%%%%%%%%%%%%%%

\end{document}